\documentclass[12pt,a4paper]{article}%
\usepackage{amssymb}
\usepackage{geometry}
\usepackage{amsmath}
\usepackage{amsfonts}
\usepackage{graphicx}%
\providecommand{\U}[1]{\protect\rule{.1in}{.1in}}
\begin{document}

\title{Is it possible to accommodate massive photons in the framework of a
gauge-invariant electrodynamics?}
\author{M. V. S. Fonseca\thanks{marcusfo@cbpf.br}, A. A. Vargas-Paredes\thanks{alfredov@cbpf.br}\\\emph{Centro Brasileiro de Pesquisas F\'{\i}sicas, Rio de Janeiro, RJ.}}
\maketitle

\begin{abstract}
The construction of an alternative electromagnetic theory that preserves
Lorentz and gauge symmetries, is considered. We start off by building up
Maxwell electrodynamics in (3+1)D from the assumption that the associated
Lagrangian is a gauge-invariant functional that depends on the electron and
photon fields and their first derivatives only. In this scenario, as
well-known, it is not possible to set up a Lorentz invariant gauge theory
containing a massive photon. We show nevertheless that there exist two
radically different electrodynamics, namely, the Chern-Simons and the Podolsky
formulations, in which this problem can be overcome. The former is only valid
in odd space-time dimensions, while the latter requires the presence of
higher-order derivatives of the gauge field in the Lagrangian. This theory,
usually known as Podolsky electrodynamics, is simultaneously gauge and Lorentz
invariant; in addition, it contains a massive photon. Therefore, a massive
photon, unlike the popular belief, can be adequately accommodated within the
context of a gauge-invariant electrodynamics.

\end{abstract}

\section{Introduction}

\qquad Maxwell electrodynamics, or its quantum version, i.e., QED, is widely
recognized as the adequate theory for the description of the electromagnetic
phenomena, because of the astonishing agreement between theory and experiment.
However, it only enjoyed this high status after some of its intrinsic problems
were solved. Among them, the most remarkable one is certainly the presence of
divergences or infinities, even at the classical level \cite{Jackson}.

This aspect of the Maxwell electromagnetic theory naturally emerges when the
self-energy of an elementary (charged) particle, like the electron, for
example, is considered. An object of this sort has no internal structure,
which means that it must be regarded (classically) as a geometric point. Its
Coulomb energy, given by%
\begin{equation}
E_{Coul}~\alpha~\int_{0}^{\infty}E^{2}dV~,
\end{equation}
where $E$ is the electron electric field, like the associated self-energy, diverges.

Objects having finite extension, on the other hand, such as composite
particles, must be described by internal degrees of freedom since in this case
the aforementioned problem, at least in principle, does not occur. Hadronic
particles, for instance, belong to this category since their static
properties, like mass, are finite and, in principle, obtained through the
quark dynamics.

In spite of the mentioned success of the electromagnetic theory, it remain
some intriguing questions that cannot be completely answered by a simple
comparison between\ experiment and theory. One of most remarkable, among
others, is the question of the massless character of the photon. From a
theoretical point of view the existence of massive photons is perfectly
compatible with the general principles of elementary particle physics. This
possibility cannot be discarded either from an experimental viewpoint. Indeed,
despite the fact that a very small value for the photon mass has not been
found experimentally up to now, this does not allows to conclude that its mass
must be identically zero. In fact, the more accurate experiments currently
avaliable can only set up upper bounds on the photon mass. Incidentally, the
recently recommended limit published by \textit{Particle Data Group} is
$m_{\gamma}\leq2\times10^{-25}GeV$ \cite{Limite superior}. On the other hand,
using the uncertainty principle, we obtain an upper limit on the photon rest
mass equal to $10^{-34}GeV$, which is found by assuming that the universe is
$10^{10}$ years old \cite{Massa do foton}. Nonetheless, the relevant question,
from a theoretical point of view, is that a nonvanishing value for the photon
mass is incompatible with Maxwell electrodynamics.

So, we can ask ourselves whether or not it would be possible to construct a
gauge-invariant electrodynamics, such as Maxwell one, but in which a massive
photon could be accommodated. At first sight, it seems that the Proca theory
\cite{Proca}, described by the Lagrangian
\begin{equation}
\mathcal{L=-}\frac{1}{4}F^{\mu\nu}F_{\mu\nu}+\frac{1}{2}m^{2}A_{\mu}A^{\mu},
\end{equation}
where$\ \ ~F^{\mu\nu}=\partial^{\mu}A^{\nu}-\partial^{\nu}A^{\mu},$ fulfills
the aforementioned requirements. Lagrangian (2) leads to massive dispersion
relations for the gauge boson, implying in a Yukawa potential in the static
case. Since this potential has a finite range, the electron self-energy is
finite \cite{Jackson}. Besides, Proca electrodynamics is Lorentz invariant.
However, gauge invariance is lost, which is certainly undesirable since, as a
consequence, this model would be in disagreement with the predictions of the
Standard Model $SU\left(  3\right)  \times SU\left(  2\right)  \times U\left(
1\right)  $ \cite{Cheng Lee}.

Other alternative models, such as the Chern-Simons \cite{Chern} and the
Podolsky \cite{Podolsky} ones, can be constructed in the same vein.

In the Chern-Simons electrodynamics a coupling between the gauge field and the
field strength is introduced into the Lagrangian through the Levi-Civita
tensor. This coupling yelds a massive dispersion relation for the gauge field.
As a result of this mechanism, a massive photon is generated. Nevertheless,
the mentioned mechanism explicitly breaks the Lorentz invariance in four
dimensions, unless a 2-form gauge field is also introduced that mixes up with
the Maxwell potential. In odd dimensions, however, this model is
simultaneously Lorentz and gauge invariant.

Podolsky electrodynamics, on the other hand, seems more interesting in
comparison to the above cited models since it can accomaddate a massive photon
without violating the Lorentz and gauge symmetries in (3+1)D.

There are other interesting aspects of Podolsky theory that deserves to be
exploited. For instance, within its context magnetic monopoles and massive
photons can coexist without conflict. That is not the case as far as the Proca
model \cite{Monopolo Proca} is concerned.

The aim of this paper is precisely to discuss the issue of the photon mass in
the framework of some outstanding electromagnetic theories. To start off,
Maxwell theory is considered in section II. In particular, it is shown in this
section that this theory can be built via simple and general assumptions; it
is also demonstrated that Lorentz and gauge invariance constrain the photon
mass to be equal to zero. In section III, we discuss the Chern-Simons theory
and prove that in odd dimensions the photon can acquire mass without breaking
the Lorentz and gauge symmetries. In section IV the Podolsky electromagnetic
theory is analyzed. We show that within the context of this model, massive
photons are allowed while the Lorentz and gauge symmetries are preserved. It
is worth mentioning that the approach to the Podolsky model we have taken in
this paper may be regarded as an alternative method to those employed by A.
Accioly \cite{Accioly} and H. Torres-Silva \cite{Torres}.

\section{Maxwell Electrodynamics}

\qquad We shall construct Maxwell electrodynamics based on the following three assumptions:

(i)Lorentz invariance holds.

(ii)There exists a Lagrangian ${\mathcal{L}}$ for the theory which is a
functional of the electron and photon fields, as well as of their first
derivatives, namely,%

\begin{equation}
\mathcal{L}=\mathcal{L}\left(  \psi,\partial_{\mu}\psi;A_{\mu},\partial_{\mu
}A_{\nu}\right)  . \label{Lagrangeano}%
\end{equation}

(iii)${\mathcal{L}}$ is invariant under a local gauge transformation.

In this spirit, we consider the following gauge transformations with respect,
respectively, to the bosonic field%
\begin{equation}
A_{\mu}~\rightarrow~A_{\mu}^{\prime}=A_{\mu}+\partial_{\mu}\beta\left(
x\right)  ,~\ \ \ \delta A_{\mu}=\partial_{\mu}\beta, \label{gauge1}%
\end{equation}
and the matter field
\begin{equation}
\psi~\rightarrow~\psi^{\prime}=\exp\left(  ie\beta\right)  \psi,~\ \ \ \delta
\psi=ie\beta\psi. \label{gauge2}%
\end{equation}
In the above equations $\beta$ is a local gauge parameter, $\beta=\beta\left(
x\right)  $.

The requirement of the invariance of the Lagrangian with respect to these
transformations, $\delta_{gauge}\mathcal{L}=0$, yields%
\begin{align}
&  \frac{\partial\mathcal{L}}{\partial\psi}\left(  ie\psi\right)  \beta
+\frac{\partial\mathcal{L}}{\partial\left(  \partial_{\mu}\psi\right)
}\left(  ie\psi\right)  \partial_{\mu}\beta+\frac{\partial\mathcal{L}%
}{\partial\left(  \partial_{\mu}\psi\right)  }\left(  ie\partial_{\mu}%
\psi\right)  \beta\nonumber\\
&  +\frac{\partial\mathcal{L}}{\partial A_{\mu}}\left(  \partial_{\mu}%
\beta\right)  +\frac{\partial\mathcal{L}}{\partial\left(  \partial_{\mu}%
A_{\nu}\right)  }\left(  \partial_{\mu}\partial_{\nu}\beta\right)  =0.
\end{align}

Now, since $\beta$ is an arbitrary parameter, we promptly obtain
\begin{equation}
\frac{\partial\mathcal{L}}{\partial\psi}\left(  ie\psi\right)  +\frac
{\partial\mathcal{L}}{\partial\left(  \partial_{\mu}\psi\right)  }\left(
ie\partial_{\mu}\psi\right)  =0.
\end{equation}
Using the Euler-Lagrange equations for the $\psi$ field in the above
expression, we then find%
\begin{equation}
\partial_{\mu}\left[  ie\frac{\partial\mathcal{L}}{\partial\left(
\partial_{\mu}\psi\right)  }\psi\right]  =0.~
\end{equation}
This result clearly shows there exists a Noetherian vector current associated
to the gauge symmetry%

\begin{equation}
j^{\mu}\equiv\frac{\partial\mathcal{L}}{\partial\left(  \partial_{\mu}%
\psi\right)  }\left(  ie\psi\right)  , \label{corrente conservada}%
\end{equation}
which is conserved ($\partial_{\mu}j^{\mu}=0).$

On the other hand, to first-order in $\beta$ derivatives, we have%
\begin{equation}
\left[  \frac{\partial\mathcal{L}}{\partial\left(  \partial_{\mu}\psi\right)
}ie\psi\right]  +\frac{\partial\mathcal{L}}{\partial A_{\mu}}=0,
\end{equation}
which can be written as%
\begin{equation}
\frac{\partial\mathcal{L}}{\partial A_{\mu}}=-j^{\mu}.
\label{acoplamento corrente}%
\end{equation}

This relation tells us how the gauge field must be coupled to a conserved
current in the Lagrangian.

Finally, the second-order derivative terms in the gauge parameter yield the
condition%
\begin{equation}
\left(  \partial_{\mu}\partial_{\nu}\beta\right)  \frac{\partial\mathcal{L}%
}{\partial\left(  \partial_{\mu}A_{\nu}\right)  }=0, \label{relacao simetrica}%
\end{equation}
which implies that the symmetric part of the derivative term in the Lagrangian
must be null, i.e,%
\begin{equation}
\frac{\partial\mathcal{L}}{\partial\left[  \partial_{(\mu}A_{\nu)}\right]
}=0. \label{parte simetrica}%
\end{equation}

Thus, we can write%
\begin{equation}
\frac{\partial\mathcal{L}}{\partial\left[  \partial_{\lbrack\mu}A_{\nu
]}\right]  }=H_{\mu\nu}, \label{antissimetrico}%
\end{equation}
where $H_{\mu\nu}$ is a totally antisymmetric rank-two tensor. Here%
\begin{equation}
\partial_{\lbrack\mu}A_{\nu]}\equiv\partial_{\mu}A_{\nu}-\partial_{\nu}A_{\mu
},
\end{equation}%
\begin{equation}
\partial_{(\mu}A_{\nu)}\equiv\partial_{\mu}A_{\nu}+\partial_{\nu}A_{\mu}.
\end{equation}

Consequently, the bosonic sector of the Lagrangian is given by%
\begin{equation}
\mathcal{L}=a\partial_{\lbrack\mu}A_{\nu]}H^{\mu\nu}+bj_{\mu}A^{\mu}.
\label{Maxwell}%
\end{equation}

The first term in Eq. (\ref{Maxwell}) is related to the vector field only, and
must be bilinear in $A_{\mu}$. As a consequence, one of the Lorentz indices of
$H_{\mu\nu}$ must necessarily be associated to the gauge field. The simplest
choice for the kinetic term, which is quadratic in $\partial_{\mu}A_{\nu}$, is%
\begin{equation}
H_{\mu\nu}=\partial_{\lbrack\mu}A_{\nu]},
\end{equation}
i.e., the tensor $H_{\mu\nu}$ can be identified with the usual electromagnetic
field strength $F_{\mu\nu}$. Taking this into account, the corresponding
Lagrangian can be written in the general form
\begin{equation}
\mathcal{L=}aF_{\mu\nu}F^{\mu\nu}+bj^{\mu}A_{\mu},
\end{equation}
where $a$ and $b$ are arbitrary constants. By analyzing the equations of
motion related to (17), it is trivial to see that a convenient choice for
these constants is $a=-\frac{1}{4}$ and $b=1$, which allows us to write%
\begin{equation}
\mathcal{L=-}\frac{1}{4}F_{\mu\nu}F^{\mu\nu}+j^{\mu}A_{\mu},
\end{equation}
which is nothing but Maxwell Lagrangian.

The field $A_{\mu}$ in (18) is massless. This raises the interesting question:
Could we have chosen the tensor $H_{\mu\nu}$ such that it contained a
gauge-invariant mass term related to $A^{\mu}$, besides the massless term?
Since in the selection of the early $H_{\mu\nu}$ we have excluded the
possibility that $A^{\mu}$ could be massive, this is a pertinent question. Let
us then discuss this possibility.

The kinetic part of the gauge field in the Lagrangian, as commented above,
must have the general form%
\begin{equation}
\mathcal{L}~\alpha~\partial_{\lbrack\mu}A_{\nu]}H^{\mu\nu}.
\end{equation}
In other words, $H_{\mu\nu}$ must be a function of $A_{\mu}$ and its first
derivatives only. Therefore, $H^{\mu\nu}$ can be written in the alternative
form%
\begin{equation}
H^{\mu\nu}=F^{\mu\nu}+h^{\mu\nu}~,
\end{equation}
where, obviously, $h^{\mu\nu}$ is an antisymmetric tensor. Accordingly,%

\begin{equation}
h^{\mu\nu}=\varepsilon^{\mu\nu\alpha\beta}\left(  ?\right)  _{\alpha}A_{\beta
},
\end{equation}
where $\varepsilon^{\mu\nu\alpha\beta}$ is the Levi-Civita tensor and the
quantity $\left(  ?\right)  _{\alpha}$ is a Lorentz vector to be determined.
There are two possibilities to be considered. The first one is to assume that
the mentioned quantity is a constant vector, which implies that it would play
the role of a fundamental quantity of nature. In this case, the aforementioned
constant vector would single out a special direction in space-time leading, as
a consequence, to a breaking of the Lorentz symmetry. The remaining choice is
$\left(  ?\right)  _{\alpha}=\partial_{\alpha}$, which would imply that the
searched quantity should be proportional to the electromagnetic
field-strength, $h^{\mu\nu}~\alpha~F^{\mu\nu}$. Thus, we come to the
conclusion that the gauge field is massless due to the two very general
assumptions considered in the construction of the Lagrangian, in addition to
the Lorentz invariance.

\section{Chern-Simons Electrodynamics}

\qquad In the preceding section we concluded that a Lagrangian which is a
functional of the electron and photon fields, as well as of their first
derivatives and, besides, is invariant under local gauge transformations and
consistent with the Lorentz symmetry, confers a massless character to the
vector field. Our proof, however, relied upon the fact that the space-time was
endowed with (3 + 1) dimensions. Yet, it is possible to show that in odd
dimensional space-times the form of the antisymmetric tensor $H_{\mu\nu}$ need
not be proportional to $F_{\mu\nu}$ only. That is the case of the so-called
Chern-Simons electrodynamics. In order to obtain the Lagrangian corresponding
to this theory we suppose that the same assumptions utilized in the
construction of the Maxwell theory still hold. As long as the quantity
$H_{\mu\nu}$ is concerned, we consider another alternative: the space-time has
(2+1)dimensions (in particular). In such a case we have to construct an
antisymmetric tensor ($h^{\mu\nu}$). This quantity can now be expressed as
follows%
\begin{equation}
h^{\mu\nu}=\varepsilon^{\mu\nu\alpha}A_{\alpha}.
\end{equation}
A Lorentz invariant term can be then constructed by contracting this term with
the usual electromagnetic tensor $F^{\mu\nu}.$ This means that the Lagrangian
can be written in the form%
\begin{equation}
\mathcal{L}~\alpha~aF_{\mu\nu}F^{\mu\nu}+b\varepsilon^{\mu\nu\alpha}F_{\mu\nu
}A_{\alpha},
\end{equation}
where $a$ and $b$ are arbitrary constants. Here $b$ has dimension of mass.

The above result may be extended to any odd dimension, because we can always
construct the Chern-Simons term through the contraction of a field with $n$
Lorentz indices with its field-strength containing $n+1$ indices. The
Levi-Civita tensor, on the other hand, will have $2n+1$ indices. For instance,
in a 5-dimensional space time, we have%
\begin{equation}
\text{\textit{Chern-Simons term}}=~\varepsilon^{\mu\nu\lambda\alpha\beta
}B_{\mu\nu}H_{\lambda\alpha\beta},
\end{equation}
with%
\begin{equation}
H_{\lambda\alpha\beta}=\partial_{\lambda}B_{\alpha\beta}+\partial_{\alpha
}B_{\beta\lambda}+\partial_{\beta}B_{\lambda\alpha}.
\end{equation}

We remark that we have only considered gauge $1-$forms to build the
Chern-Simons term; nevertheless, it is also possible to use a gauge $2-$form
(the so called$\ $"$BF~$" term $\left(  \varepsilon^{\mu\nu\kappa\lambda
}B_{\mu\nu}F_{\kappa\lambda}\right)  $) to accomplish this goal. However, in
order to avoid the introduction of new degrees of freedom \cite{Sorella}, we
have opted in this paper to work in the Chern-Simons scenario.

\section{Podolsky Electrodynamics}

\qquad In the preceding sections, we have found that in $(1+3)D$ the vector
gauge field is massless as a consequence of the very general assumptions made
in order to build the associated Lagrangian. That is not the case whenever odd
dimensional space-times are concerned. Indeed, in these space-times a mass
term for the vector field is allowed. We are now ready to focus on the issue
theme of this work, i.e., the question of whether or not massive photons can
be accommodated in the context of a gauge-invariant electromagnetic theory in
$(3+1)D$. To do that, we shall relax one of the assumptions made in the
construction of the preceding electrodynamics, namely, the one that forbids
the presence of higher derivatives of the gauge field in the Lagrangian. As a
result, the gauge sector will be altered while the matter contribution remains
unchanged. To be more explicit, let us suppose that the Lagrangian is as
follows%
\begin{equation}
\mathcal{L}=\mathcal{L}\left(  \psi,\partial_{\mu}\psi;A_{\mu},\partial_{\nu
}A_{\mu},\partial_{\lambda}\partial_{\mu}A_{\nu}\right)  .
\end{equation}
Imposing now that (26) is invariant with respect to the transformations
(\ref{gauge1}) and (\ref{gauge2}) yields%
\begin{align}
&  \frac{\partial\mathcal{L}}{\partial\psi}\left(  ie\beta\psi\right)
+\frac{\partial\mathcal{L}}{\partial\left(  \partial_{\mu}\psi\right)
}\left[  ie\left(  \partial_{\mu}\beta\right)  \psi\right]  +\frac
{\partial\mathcal{L}}{\partial\left(  \partial_{\mu}\psi\right)  }\left(
ie\beta\partial_{\mu}\psi\right)  +\frac{\partial\mathcal{L}}{\partial A_{\mu
}}\left(  \partial_{\mu}\beta\right)  \nonumber\\
&  +\frac{\partial\mathcal{L}}{\partial\left(  \partial_{\mu}A_{\nu}\right)
}\left(  \partial_{\mu}\partial_{\nu}\beta\right)  +\frac{\partial\mathcal{L}%
}{\partial\left(  \partial_{\lambda}\partial_{\mu}A_{\nu}\right)  }\left(
\partial_{\lambda}\partial_{\mu}\partial_{\nu}\beta\right)  =0.
\end{align}

Noting, as it was expected, that the lower-order terms in the gauge parameter
$\beta$ have not changed, we come to the conclusion that the conditions
(\ref{corrente conservada}), (\ref{acoplamento corrente}) and
(\ref{antissimetrico}) will not be altered. The term with third order
derivatives, on the other hand, tells us that
\begin{equation}
\frac{\partial\mathcal{L}}{\partial\left(  \partial_{\mu}\partial_{\nu
}A_{\lambda}\right)  }\left(  \partial_{\lambda}\partial_{\mu}\partial_{\nu
}\beta\right)  =0.
\end{equation}

A possible solution to (27), is%
\[
\left(  \partial^{\lambda}F^{\mu\nu}\right)  G_{\lambda\mu\nu},
\]
where the quantity $G$ cannot be symmetric with respect to all its indices due
to the Lorentz invariance. Actually, $G_{\mu\nu\lambda}$ must be antisymmetric
in the last two indices, i.e., $G_{\mu\nu\lambda}=G_{\mu\left[  \nu
\lambda\right]  }$ so that we may identify $\frac{\partial\mathcal{L}%
}{\partial\left(  \partial_{\lambda}\partial_{\lbrack\mu}A_{\nu]}\right)  }$
with $G_{\lambda\left[  \mu\nu\right]  }$. As a consequence,%

\begin{equation}
\frac{\partial\mathcal{L}}{\partial\left(  \partial_{\lambda}\partial
_{\lbrack\mu}A_{\nu]}\right)  }=G_{\lambda\left[  \mu\nu\right]  },
\end{equation}
where%

\[
G_{\lambda\left[  \mu\nu\right]  }=G_{\lambda\mu\nu}-G_{\lambda\nu\mu}.
\]

Therefore, the corresponding Lagrangian must have the general form%
\begin{equation}
\mathcal{L}=aF^{\mu\nu}F_{\mu\nu}+b\left(  \partial^{\lambda}F^{\mu\nu
}\right)  G_{\lambda\left[  \mu\nu\right]  }+cj^{\mu}A_{\mu}.
\end{equation}

The functional above is a function of $A_{\mu},~\partial_{\mu}A_{\nu}$ and
$\partial_{\mu}\partial_{\nu}A_{\lambda}$. Now, since the second derivatives,
$\partial_{\mu}\partial_{\nu}$, commute, the antisymmetric part of
$G_{\lambda\left[  \mu\nu\right]  }$ must be constructed with first
derivatives of the field $A_{\mu}$ only. Since the term $\left(
\partial^{\lambda}F^{\mu\nu}\right)  G_{\lambda\left[  \mu\nu\right]  }$ must
be quadratic in the gauge field, the remaining index of $G_{\lambda\left[
\mu\nu\right]  }$ will be identified with the first derivative of the
antisymmetric part of the aforementioned tensor. This means that the quantity
$G_{\lambda\left[  \mu\nu\right]  }$ is nothing but the derivative of the
usual field-strength tensor. Hence, the Lagrangian is given by%
\begin{equation}
\mathcal{L}=-\frac{1}{4}F^{\mu\nu}F_{\mu\nu}+\frac{b^{2}}{4}\left(
\partial^{\lambda}F^{\mu\nu}\right)  \partial_{\lambda}F_{\mu\nu}+j^{\mu
}A_{\mu},\label{dispersion relation}%
\end{equation}
where judicious values for the arbitrary constants were chosen. The above
Lagrangian is known as the Podolsky Lagrangian. Here $b$ is a constant with
dimension of $\left(  mass\right)  ^{-1}$.

Now, in order not to conflict with well-established results of QED, the
parameter $b$ must be very small, which implies that the massive photon,
unlike what is claimed in the literature, is a heavy photon. Indeed, recently,
Accioly and Scatena \cite{Scatema} found that its mass is $\sim42GeV$, which
is of the same order of magnitude as the mass of the $W\left(  Z\right)  $
boson \cite{Amsler}. This is an interesting coincidence.

To conclude, we call attention to the fact that Podolsky theory plays a
fundamental role in the discussion about the issue of the compatibility
between magnetic monopoles and massive photons.

\section*{Acknowledgments}

We are grateful to O. A. Battistel, A. Accioly and J. Helay\"{e}l-Netto for
helpful discussions and the reading of our manuscript. CNPq-Brazil is also
acknowledged for the financial support.

\end{document}